# Sécurité dans les SI & *social engineering* - un état des lieux

**Florence Sèdes[1], Jonathan Degrace[2]**

1. sedes@irit.fr
IRIT - Université Toulouse 3 Paul Sabatier
118 Route de Narbonne
31062 Toulouse cedex 9
2. jonathan.degrace@medes.fr
MEDES - Clinique Spatiale
2 avenue de l'aérodrome de Montaudran
31 400 Toulouse

RESUME. *De grandes transformations liées aux technologies de l'information touchent les Systèmes d'Information (SI) qui soutiennent les processus métier des organisations ainsi que leurs acteurs. Le déploiement dans un environnement complexe concernant des données sensibles, massives et hétérogènes génère des risques aux impacts juridiques, soci(ét)aux et financiers. Ce contexte de transition et d'ouverture rend la sécurité de ces SI centrale dans les préoccupations des organisations. La numérisation de tous les processus et l'ouverture aux dispositifs IdO (Internet des Objets) a favorisé l'apparition d'une nouvelle forme de criminalité : la cybercriminalité.*
*Ce terme recouvre des actions «malicieuses» (malveillantes) dont la majorité sont désormais perpétrées à l'aide de stratégies de social engineering, phénomène permettant une exploitation combinée des vulnérabilités «humaines» et des outils numériques. La «maliciosité» de telles attaques réside dans le fait qu'elles transforment les utilisateurs en facilitateurs des cyberattaques, au point d'en être perçus comme le «maillon faible» de la cybersécurité. Les politiques de déploiement s'avérant insuffisantes, il est nécessaire de réfléchir à des étapes amont : savoir anticiper, analyser signaux faibles et outliers, détecter précocement et réagir promptement sont des questions prioritaires nécessitant une approche axée sur la prévention et la coopération. Dans cet état des lieux, nous proposons un travail de synthèse de la littérature et des pratiques professionnelles à ce sujet.*

ABSTRACT. *Major transformations related to information technologies affect Information Systems (IS) that support the business processes of organizations and their actors. Deployment in a complex environment involving sensitive, massive and heterogeneous data generates risks with legal, social and financial impacts. This context of transition and openness makes the* *security of these IS central to the concerns of organizations. The digitization of all processes and the opening to IoT devices (Internet of Things) has fostered the emergence of a new form of crime, i.e. cybercrime.*

*This generic term covers a number of malicious acts, the majority of which are now perpetrated using social engineering strategies, a phenomenon enabling a combined exploitation of «human» vulnerabilities and digital tools. The maliciousness of such attacks lies in the fact that they turn users into facilitators of cyber-attacks, to the point of being perceived as the «weak link» of cybersecurity.*

*As deployment policies prove insufficient, it is necessary to think about upstream steps: knowing how to anticipate, identifying weak signals and outliers, detect early and react quickly to computer crime are therefore priority issues requiring a prevention and cooperation approach.*

*In this overview, we propose a synthesis of literature and professional practices on this subject.*

*Mots-clés* : *social engineering - cybercriminalité - attaques - prévention.*
K<sup>EYWORDS</sup>: *social engineering - cybercrime - attacks – prevention.*

1. Introduction

De grandes transformations liées aux technologies de l'information touchent les Systèmes d'Information (SI) qui soutiennent les processus métier des organisations ainsi que leurs acteurs. Le déploiement dans un environnement complexe concernant des données sensibles, massives et hétérogènes génère des risques aux impacts juridiques, sociaux, et financiers. Ce contexte de transition et d'ouverture rend la sécurité de ces SI centrale dans les préoccupations des organisations. La numérisation de tous les processus et l'ouverture aux dispositifs IdO (Internet des Objets / Internet of Things (IoT)) a favorisé l'apparition d'une nouvelle forme de criminalité : la cybercriminalité.

Ce terme générique recouvre un certain nombre d'actes « malicieux » (malveillants) dont la majorité sont désormais perpétrés à l'aide de stratégies de *social engineering* (le terme d'ingénierie sociale en français ne recouvre pas les mêmes acceptions), phénomène permettant une exploitation combinée des vulnérabilités « humaines » et des outils numériques. La « maliciosité » de telles attaques réside dans le fait qu'elles transforment les utilisateurs en facilitateurs des cyberattaques, au point d'en être perçus comme le « maillon faible » de la cybersécurité.

Les particuliers, les entreprises, les institutions et les États sont confrontés au défi de trouver une réponse à ces atteintes. Néanmoins, les moyens juridiques, techniques, économiques et culturels mis en place sont encore insuffisants : loin d'être éradiquée, l'utilisation du *social engineering* à des fins illicites poursuit son essor. Les éléments factuels illustrent ce constat : [1] relève que 74 % des failles de sécurité incluent l'élément humain, que 50 % des attaques de *social engineering* sont de la compromission d'email professionnel et que la cible numéro deux des cyberattaques sont les individus.

Les domaines liés à ces problématiques, comme les aspects juridiques du *social engineering* et les sanctions encourues, ou le rôle des États et des organisations internationales dans la lutte contre la cybercriminalité ne seront pas envisagés dans cette synthèse. L'objectif de ce papier est de situer l'état des lieux des connaissances actuelles concernant le fonctionnement et la prévention du *social engineering* :

- À quel point est-il impliqué dans les incidents en cybersécurité ?
- Quels sont ses vecteurs d'attaques ? Comment fonctionnent-ils ?
- Existe-t-il des mesures de prévention efficaces ?

2. Définition du *social engineering*

Historiquement, le terme de *social engineering* vient des sciences politiques. Apparu sous la plume de l'économiste britannique John Gray en 1842, dans son ouvrage, "An efficient Remedy for the distress of Nations", il désignait à l'origine les experts en charge des questions politiques et sociales, par exemple la résolution de la grande famine irlandaise (1845-1852) qui allait sévir quelques années plus tard. Il sera plus tard usité en sciences sociales, e.g. par J. M. Hatfield 2018 [22] et Slade, John A. 1929 [23]. Son entrée dans le monde cyber se fait dans les années 50 avec le phénomène des "phone phreaking" et les opératrices manuelles, dans le but de détourner le fonctionnement des lignes téléphoniques afin d'accéder à des services spéciaux et / ou de ne pas payer les communications.

Le *social engineering*, encore appelé fraude psychologique ou piratage psychologique, désigne l'utilisation des techniques d'escroqueries historiques adaptées au monde du numérique : on y retrouve des techniques comme la recherche d'informations sur la cible, le mensonge, l'utilisation d'une fausse identité, la tromperie ou la manipulation.

Le but est d'acquérir la confiance de l'autre afin de lui faire réaliser une action frauduleuse comme l'accès à des informations, des biens, des services ou des lieux sécurisés (voir un exemple historique avec le Tupolev, "cousin" du Concorde, "Tupolev-Tu-144-l-espionnage-industriel-au-cœur-de-la-guerre-froide" [15]).

Dans le cadre de cet état de l'art, notre focus se fera sur le *social engineering* impliqué dans les cyberattaques.

Comme évoqué précédemment, 74 % des violations de données incluent un élément humain : vol d'identité, mésusage des accès à privilèges, mauvaises configurations ou *social engineering*, les classant dans le top 3 des causes de violations de données dans 8 secteurs sur 9, et numéro 2 sur la scène internationale [1].

Dans les seules attaques de *social engineering*, la compromission d'email professionnel, à l'aide de technique de phishing, a augmenté de près de 50 % entre 2022 et 2023. Le *phishing*, et ses dérivés (*spear-phishing, vishing, smishing,* etc. cf. 3.1 ci-après), représentent à eux seuls 44 % des attaques en 2023, pour un coût médian de 50 000$ par attaque. C'est actuellement le principal vecteur d'attaque. Parmi ces attaques, 83 % sont des acteurs extérieurs à l'organisation, dont le but est purement lucratif.

À la vue des chiffres indiqués par le rapport 2023 de Verizon [1], on peut ne peut que s'accorder sur l'importance du facteur humain dans la prévention des risques en cybersécurité. En effet, les *social engineers* ne font que profiter de certains des fonctionnements de l'être humain, couplés à l'utilisation des moyens numériques et des techniques de manipulation, comme effet de levier pour leurs attaques (Wang et al. 2021 [2]).

### 3. Les attaques par *social engineering*

*3.1 Différentes formes d'attaques*

Les attaques peuvent employer, ou non, des moyens techniques comme les emails, les sms, les appels, etc. [30].

Les attaques les plus courantes sont le *phishing* et le *spear phishing* par mail [1]. Le *spear phishing* est un *phishing* personnalisé pour la cible. Nous trouvons certains de ses "dérivés" dans les *smishing* (*phishing* via SMS) ou le *vishing* (*phishing* vocal via message audio ou appel direct). Ces attaques entrent dans la catégorie du *harpooning*, ou hameçonnage en français. Elles consistent à envoyer un message, sous un faux prétexte, pour tromper les victimes afin de leur faire commettre une action qui leur sera préjudiciable, tel un mail urgent de l'école de leurs enfants par exemple, qui leur fait ouvrir un document qui cacherait un ransomware *(phishing / spear-phishing)*. L'envoi d'un sms incitant à ouvrir un lien menant à un faux site de leur banque dans le but d'en dérober les identifiants et mot de passe *(smishing)*, ou un appel, suite à de prétendus prélèvements frauduleux, leur demandant de remettre leurs cartes avec les codes confidentiels à un pseudo-coursier prétendument mandaté par le conseiller bancaire *(vishing)* constituent des stratégies bien rodées. De telles attaques peuvent également avoir lieu à partir d'autres supports comme les réseaux sociaux [10] [24] [31].

À cela, nous pouvons ajouter les attaques physiques en face à face ou à l'aide de *tailgating*, technique consistant à "coller" quelqu'un afin de passer un portique sécurisé par badge par exemple. Un attaquant déguisé en employé pourrait utiliser cette technique dans le but de pénétrer un bâtiment avant de se fondre dans la masse et d'essayer de mener à bien ses actes illicites. Party et Rajendran 2019 [11] présentent le *shoulder surfing* qui consiste à observer par-dessus l'épaule de la victime quand cette dernière saisit des données. On peut facilement imaginer profiter d'un voyage en train ou d'une séance de travail à la terrasse d'un café de notre victime, pour lire ou photographier un écran à l'insu de son propriétaire. La fouille des poubelles de notre cible reste aussi une forme d'attaque de *social engineering*, que cela soit pour trouver des informations pour une attaque ciblée (*spear-phishing*) ou récupérer sur des supports non détruits des informations, pour elle/lui sans importance, mais qui peuvent s'avérer précieuses (dates anniversaires, nom d'animaux, etc. souvent à la base de la création de passwords).

Il existe également des attaques plus techniques et moins directes, comme l'attaque du point d'eau. Cette dernière consiste en la compromission d'un point de rencontre (ex: site web) fréquemment visité par les victimes, comme peuvent parfois le faire des prédateurs avec leurs proies. Elle permet d'atteindre des victimes de façon indirecte par rebond lorsque l'attaque directe est trop complexe ou risquée (S. Kaushalya, R. Randeniya, and A. Liyanage, 2018 [12]).

L'Open Source INTelligence (OSINT) est régulièrement utilisée dans la préparation des attaques (Wang et Al, 2020 [13]). L'OSINT consiste en la recherche d'information à partir de sources ouvertes légalement accessibles, sur internet ou ailleurs. Ces informations peuvent provenir des réseaux sociaux, des registres légaux ou encore de différents médias.

L'arrivée des différentes formes de l'Intelligence Artificielle (IA) et du Machine Learning (ML) permet d'amplifier les effets des attaques de *social engineering*. Ainsi, on assiste à une quasi-industrialisation de la recherche d'informations et des attaques personnalisées de type *spear phishing*. Nous assistons également à l'arrivée d'IA "d'attaque" capables de s'autocorriger afin de s'adapter de plus en plus finement à leurs interlocuteurs pour les tromper (Mouton et al. 2014 [16], Schmitt et Flechais 2023 [17]).

Une liste non exhaustive de différentes méthodes d'attaques peut être trouvée dans les publications de Yasin et al. 2019 [10], Party et Rajendran 2019 [11] et S. Kaushalya, R. Randeniya, and A. Liyanage, 2018 [12] et infosecawerness [28].

### 3.2 Quelles sont les failles humaines et comment sont-elles utilisées ?

Pour la réussite des différentes techniques énoncées, les *social engineers* s'appuient sur différents aspects du fonctionnement humain, parmi lesquels nous retrouvons certains mécanismes cognitifs, les biais des cibles, les besoins sociaux, les stéréotypes, les heuristiques de pensées et les émotions des individus (Wang et al 2021 [2], Laurent Bègue et Olivier Desrichard 2013 [14] et Yasin et al. 2019 [10]).

Les *social engineers* ne sont bien évidemment eux-mêmes pas à l'abri de ces mêmes biais et autres stéréotypes. On peut citer à ce titre le stéréotype de genre qui induit un biais qui fait que les femmes, réputées moins versées en technologie numérique, sont plus ciblées que les hommes [26].

Ils peuvent aussi compter sur différentes techniques de manipulations ou d'influence pour les aider dans leurs tâches.

Tout cela a pour effet de créer des schémas globaux de potentielles vulnérabilités exploitables en *social engineering* : par exemple, nous serons plus à même d'aider une personne plus proche ou qui nous est physiquement agréable, par biais de halo. D'autres biais cognitifs sont mobilisés, comme le biais de conformité (tendance à penser et agir comme les autres le font), le biais d'ancrage (s'en tenir au premier élément d'information entendu comme référence), l'excès de confiance (tendance à surestimer ses capacités), ... À partir de cette approche, le spécialiste en *social engineering* peut déployer un certain nombre d'outils, que cela soit lors d'une attaque physique ou « virtuelle », à l'aide de l'un des moyens de communication à sa disposition.

De surcroît, lorsque certains traits de personnalités (au sens du *big five*[1]) sont dominants, ils augmentent le risque de tomber dans le piège d'une attaque par *social engineering* [25].

---

1. https://fr.wikipedia.org/wiki/Mod%C3%A8le_des_Big_Five_(psychologie)

Il est ainsi facile d'imaginer qu'un *social engineer* joue, par exemple, sur le biais de préférence endogroupe [14] à l'aide d'une fausse identité. Ainsi, il peut se rapprocher de sa cible et lui soutirer des informations, des accès ou des biens.

Nous présentons dans le tableau ci-dessous un échantillon non exhaustif des biais, techniques d'influence et de manipulation, que pourrait rencontrer les cibles, lors d'une attaque de *social engineering*. Pour plus d'information, on peut se reporter à Wang et al 2021 [2], Yasin et al. 2019 [10], Caldini 2014 [34] et R-B Joules et J-L Beauvois 2022 [35].

| Biais (cognitif) de la cible | Fonctionnement | Conséquences éventuelles |
|---|---|---|
| Biais de confirmation d'hypothèse | Tendance à rechercher des informations qui confirment nos croyances existantes et à ignorer les informations qui les contredisent : on filtre et on trie les informations selon qu'elles correspondent à nos attentes ou non. | Risque de divulgation d'informations, d'accès non autorisé ou d'action frauduleuse. |
| Aversion à la perte | L'aversion à la perte implique que les individus sont plus sensibles aux perspectives de pertes qu'à celles associées aux gains. | Cela peut mettre la victime sous pression et lui faire prendre un risque inconsidéré dans une attaque de type "Appel du président". |
| Effet Lake Wobegon (ou biais d'auto-complaisance) | L'effet Lake Wobegon traduit la tendance (inconsciente) à penser que nous sommes bien meilleurs que nous ne le sommes en réalité. | La victime pourrait surestimer sa capacité à se défendre face aux différentes formes d'attaques. |
| **Technique d'influence** | **Fonctionnement** | **Conséquence éventuelle** |
| Sentiment de proximité par appartenance | Sentiment de proximité supposé envers un individu qui augmente les probabilités de l'aider. | Risque de divulgation d'informations, d'accès non autorisé ou d'action frauduleuse. |
| Attractivité physique ou empathique | Sentiment d'attractivité envers un individu qui augmente les probabilités de l'aider. | Risque de divulgation d'informations, d'accès non autorisé ou d'action frauduleuse. |
| Conformisme aux figures d'autorités | Céder plus facilement aux figures d'autorités par normes sociales. | Risque de divulgation d'informations, d'accès non autorisé ou d'action frauduleuse. |
| **Technique utilisée** | **Fonctionnement** | **Conséquences éventuelles** |
| Distraction | Créer une distraction modérée dans le but de surcharger mentalement et diminuer la capacité de réflexion de la cible. | Risque de divulgation d'informations, d'accès non autorisés ou d'action frauduleuse. |
| Fake ID | Utilisation d'une fausse identité sociale dans le but de paraître plus proche de la cible et obtenir son aide. | Tromperie sur l'identité de l'attaquant pouvant mener à un risque de divulgation d'informations, d'accès non autorisés ou d'action frauduleuse. |
| Usage des symboles d'autorités | Utilisation des symboles d'autorités reconnus (ex : blouse, façon de s'exprimer) pour appuyer l'utilisation du fake ID. | Comme le fake ID, avec une pression accrue pouvant mener à un risque de divulgation d'informations, d'accès non autorisés ou d'action frauduleuse plus important. |

*3.3 Mise en œuvre*

D'après Mouton et al. 2014 [16], une attaque par *social engineering* se décompose en six phases. Le nombre d'étapes et le déroulement pourront varier selon le type d'attaque (physique ou *phishing* par exemple). Nous pouvons les détailler de la manière suivante.

Phase 1 : Formulation de l'attaque :
On détermine quel est le but de cette attaque.
Phase 2 : Récolte d'information :
Analyses des sources d'informations disponibles et de leurs utilisations possibles.
Récolte de l'information
Phase 3 : Préparation de l'attaque
À l'aide des informations récoltées, on vient définir la méthode, éventuellement qui exécute l'attaque et selon quel scénario.
Phase 4 : Développement de la relation de confiance :
Début de l'attaque, tentative de mise en confiance de la cible et de consolidation de la relation.
Phase 5 : Exploitation de la relation de confiance :
On exploite la relation de confiance afin d'obtenir une action frauduleuse de la part de la cible, sans que celle-ci le réalise.
Phase 6 : Debrief
On ramène la cible à un état émotionnel positif / neutre dans le but d'éviter toute forme de culpabilisation ou de sentiments négatifs. Ces derniers pourraient entraîner un refus ultérieur ou une alerte.

Selon les vecteurs d'attaques choisis, les phases 4 et 5 peuvent être imbriquées et la phase numéro 6 peut ne pas avoir lieu. Prenons un exemple de scénario fictif.

Nous sommes les attaquants et nous voulons des informations sur l'administrateur système d'une TPE spécialisée dans l'innovation médicale afin de s'introduire dans le système d'information pour y dérober des informations confidentielles.

Lors de la surveillance et des recherches préliminaires, nous avons remarqué que l'un des employés du service logistique était un grand supporter de l'équipe de rugby de sa ville. Il assistait à tous les matchs de son équipe favorite.

Il n'est pas difficile de se rendre à un match en feignant être fraîchement arrivé en ville et vouloir s'intégrer et rencontrer les autres habitants à travers leur passion commune pour ce sport.

De discussion en discussion et en orientant un peu cette dernière lors des troisièmes mi-temps, nous récoltons de plus en plus d'informations sur l'administrateur système, adresse, horaires de travail et autres habitudes (il n'est pas trop du matin par exemple), prénom de ses quatre enfants (dont un avec un prénom rare et compliqué), goûts musicaux ou culinaires, etc.

Ainsi, il nous est facile de créer un mail de *spear phishing* sur mesure pour créer un accès au réseau de l'entreprise, mais surtout découvrir le mot de passe administrateur, qui n'était autre que le prénom rare de son enfant, de l'Active Directory, et se créer un accès discret au lieu de stockage des documents sensibles.

Par la suite, nous avons gardé le contact avec le fan de rugby jusqu'à ne plus avoir besoin de lui et disparaître dans la nature en prétextant un énième déménagement professionnel, afin de ne pas éveiller les soupçons chez lui.

La mission est réussie. Nous avons récupéré les informations nécessaires pour notre espionnage industriel, et nous n'avons pas été repérés par notre victime.

Dans cette courte fiction, nous retrouvons bien les six phases de notre attaque :
*Phase 1 :*
Le but de l'attaque : dérober des informations confidentielles.
*Phase 2 :*
Recherche d'information :
Observation des utilisateurs, recherche sur une cible potentielle,
*Phase 3 :*
Comment va-t-on avoir accès au réseau de la TPE ? Comment récolter des informations auprès du logisticien ?
*Phase 4 :*
Prise de contact et discussion lors des matches et des troisièmes mi-temps.
*Phase 5 :*
Création d'un *spear phishing*, récupération des données confidentielles.
*Phase 6 :*
Rompre le contact de manière délicate avec le logisticien. Le but : ne pas éveiller les soupçons.

L'utilisation de l'IA peut se retrouver à chacune de ces phases (Mouton & al. 2014 [16]), facilitant l'élicitation de données sur le profil des cibles, amplifiant les effets de l'attaque et surtout facilitant l'accès aux attaques par *social engineering* à beaucoup plus de personnes.

**4. Contre-mesure existante et contrepartie / efficacité**

La méta-analyse de Syafitri et al, 2022 [3], concernant la prévention du *social engineering*, semble montrer des résultats.

On constate que c'est la nouveauté de l'attaque qui crée son opportunité, et conditionne son succès : une fois largement informée, la population sait réagir et éviter la chausse-trappe, et il est alors nécessaire de renouveler la stratégie afin de surprendre à nouveau. L'apprentissage est donc un élément de la solution, donc la formation et la prévention des axes à privilégier.

La prévention contre le *social engineering* peut dès lors revêtir de multiples formes : *serious games*, entraînement par le test (faux mail de *phishing* par exemple), formations dédiées.
Devant une telle diversité, il peut être utile de connaître les performances et les impacts des différentes méthodes de prévention : niveaux d'efficacité, contextes d'utilisation, limites, afin de cibler les bons outils au bon moment dans le bon contexte.

On remarque que les actions de préventions menées paraissent principalement axées sur la sensibilisation, parfois accompagnées d'entraînements, sur des vecteurs d'attaques spécifiques (exemple : *phishing, vishing*, de plus en plus fréquemment en lien avec les évolutions multimodales de l'IA).

D'après (Mouton et al. 2015 [4]) nous avons également des processus d'aide à la détection tels que le SEADMv2 (Security Education Training Awareness) ou des propositions alliant les capacités de détection des utilisateurs seniors avec du ML (Burda 2020 [5]).

Les actions de prévention, comme la sensibilisation ou l'entraînement sur des vecteurs d'attaques spécifiques, semblent avoir des effets à long terme discutables. En effet, M. Junger et al, 2017 [9] montrent que l'augmentation des compétences de défense contre le *social engineering* n'est pas automatiquement corrélée à chaque nouvelle séance de formation.

Une étude d'Olivier de Casanove et Florence Sèdes 2022 [7] portant sur l'amélioration des programmes SETA (Security Education Training Awareness) a déjà proposé une méthodologie d'amélioration de la prévention. Elle s'appuie notamment sur l'utilisation de l'outil PDCA (Plan – Do – Check – Act) afin de créer un framework d'amélioration continue à l'aide de collecte et d'analyse des données lors des différentes campagnes de prévention et de sensibilisation.

La méta-analyse de Bullee et Junger 2020 [20] et l'étude de P. Kumaraguru [19] apportent des pistes de travail dans le cadre de la préparation d'un plan de prévention. Il faut cependant noter que les interventions sont principalement exécutées en laboratoire et sur un seul vecteur d'attaque à la fois. Néanmoins, elles respectent dans l'ensemble les principes de prévention et de formation validés par les spécialistes en science de l'éducation, aussi bien pour les adultes [32] que pour les enfants [33], ouvrant ainsi la voie à une possibilité de prévention dès le plus jeune âge.

Nous pouvons essayer d'en déduire les recommandations suivantes :

➔ Former tout le monde, pas seulement celles et ceux ayant échoué aux tests.
➔ Tester avec un feedback immédiat pour expliquer les erreurs commises.
➔ Travailler sur des sujets spécifiques.
➔ Préférer une modalité d'intervention dynamique avec composante verbale.
➔ Gamifier pour faciliter l'apprentissage.
➔ Avoir recours à des illustrations plutôt qu'une simple présentation ou du texte seul.
➔ Assimiler le mécanisme des URL qui a un effet relativement important.
➔ Eviter les messages d'avertissement, quasi-inutiles.
➔ Sensibiliser, entraîner et tester régulièrement.

Une recommandation globale et partagée consiste à prôner une approche du *social engineering* inclusive [28].

**5. Défis et orientations futures**

Comme évoqué par Schmitt et al. [17] et Falade et al. [18], la lutte contre le *social engineering* amplifié par IA et ML ne pourra se passer de la prévention humaine. Elle nécessitera également la collaboration des différents secteurs afin de développer des outils dont des IA pour venir soutenir la défense, par exemple en permettant de découvrir les variations, même très légères, laissées par les IA génératives dans leurs productions.

Malgré les investissements pour établir des stratégies efficaces contre les attaques, les méthodes de détection existantes sont limitées et les contre-mesures inefficaces pour faire face au nombre croissant d'attaques, en raison des vulnérabilités technologiques et humaines exploitées. Parce que l'humain est un défi pour la sécurité de tout réseau, il est important de développer les programmes de formation pour les employés et au-delà tout citoyen, cible potentielle, en investissant dans l'éducation à la cybersécurité :
- Quelles sont les meilleures pratiques pour sensibiliser et former les utilisateurs aux risques du *social engineering* ?
- Comment les organisations peuvent-elles collaborer pour lutter contre la cybercriminalité ?
- Quelles sont les implications éthiques de l'utilisation de l'IA dans la cybersécurité ?
- Comment garantir la protection de la vie privée et de la confiance numérique face aux menaces du *social engineering* ?

Les politiques de déploiement s'avérant insuffisantes, la nécessité de réfléchir à des étapes amont se fait jour : savoir anticiper, imaginer en se mettant « à la place » des attaquants, détecter précocement signaux faibles et *outliers* et réagir promptement face à la délinquance informatique sont alors des questions prioritaires nécessitant une approche axée sur la prévention et la coopération.

## 6. Conclusion

Dans cette synthèse, nous avons donné un aperçu des attaques de *social engineering*, des techniques de détection existantes, et des contre-mesures, à l'efficacité limitée, pour des raisons technologiques ou humaines : un système de sécurité robuste peut être facilement contourné par une simple attaque de *social engineering*. De telles attaques ont augmenté en intensité et en nombre et causent d'importants dommages émotionnels et financiers, depuis les institutions publiques jusqu'au particulier, en passant par les entreprises de toutes tailles.

Cette revue de littérature a été synthétisée pour donner un aperçu des différentes formes d'attaques existantes ainsi que des leviers humains sur lesquels elles s'appuient.

Le *social engineering*, augmenté des technologies d'IA qui se généralisent (multimodalité, vidéo, audio…) reste une menace importante en constante évolution, qui ouvre de nouvelles perspectives : peut-on réellement former efficacement sur tous les vecteurs d'attaques connues ou non ? que se passe-t-il lorsqu'un utilisateur ou une utilisatrice se retrouve face à une méthode d'attaque jusqu'alors inconnue ?

La prévention et la formation continue des utilisateurs et des utilisatrices reste un socle indispensable pour une défense efficace, en synergie avec les moyens de défense technique. Le *social engineering* est un défi majeur pour la sécurité des SI et la cybersécurité. Une approche multidimensionnelle et collaborative est nécessaire pour contrer cette menace en constante évolution, la sensibilisation, la formation et la recherche de solutions innovantes s'avérant essentielles pour protéger les individus, les organisations et les sociétés.

Des approches pluridisciplinaires seront nécessaires pour améliorer les programmes de prévention, développer de nouveaux outils et essayer de prendre de l'avance sur les cybercriminels. L'impact du *social engineering* sur la vie privée et la confiance numérique est essentiel à prendre en compte, de même que les implications éthiques de l'utilisation de l'IA dans la cybersécurité.

## 7. Références